\documentclass[12pt]{iopart}

\usepackage{iopams}

\usepackage[nodots]{numcompress}
\usepackage{verbatim}
\usepackage{graphicx}
\usepackage{fancybox}
\usepackage{subfigure}
\usepackage[usenames,dvipsnames]{color}
\usepackage{sidecap}
\usepackage{caption}
\usepackage{ifthen}
\usepackage{textcomp}
\usepackage[all]{xy}

\newcommand{\N}{{\mathbb N}}

\newcommand{\R}{{\mathbb R}}

\newcommand{\dem}{{\em Proof: \;}}
\newcommand{\fdem}{\hfill $\square$}

\newtheorem{teo}{Theorem}[section]
\newtheorem{lema}[teo]{lemma}
\newtheorem{cor}[teo]{Corollary}
\newtheorem{prop}[teo]{Proposition}

\newtheorem{defi}{Definition}[section]

\begin{document}

\title[Two-dimensional Minkowski causal automorphisms]{Two-dimensional Minkowski causal automorphisms and conformal maps}

\author{Juan Manuel Burgos}

\address{Instituto de Matem\'aticas, Universidad Nacional Aut\'onoma de M\'exico, Unidad Cuernavaca.\\ Av. Universidad s/n, Col. Lomas de
Chamilpa. Cuernavaca, Morelos M\'exico, 62209.}
\ead{burgos@matcuer.unam.mx}

\begin{abstract}
Treating the two-dimensional Minkowski space as a Wick rotated version of the complex plane, we characterize the causal automorphisms in two-dimensional Minkowski space as the M\"{a}rzke-Wheeler maps of a certain class of observers. We also characterize the differentiable causal automorphisms of this space as the Minkowski conformal maps whose restriction to the time axis belongs to the class of observers mentioned above. We answer a recently raised question about whether causal automorphisms are characterized by their wave equation. As another application of the theory, we give a proper time formula for accelerated observers which solves the twin paradox in two-dimensional Minkowski spacetime.
\end{abstract}

\vspace{2pc}
Published under minor corrections in \emph{Classical and Quantum Gravity}.

\maketitle

\section{Introduction}

In 1964, Zeeman \cite{Ze} proved the following rigidity theorem on causal automorphisms:

\begin{teo}
In $n\geq3$ dimensional Minkowski spacetime, every causal automorphism is the composite of a translation, a dilation and an orthochronous Lorentz transformation.
\end{teo}

Recently, the solution to the long standing problem of the characterization of causal automorphisms in two dimensional spacetime was given by Kim \cite{Ki} (see also \cite{Lo}). Treating the two-dimensional Minkowski space as a Wick rotated version of the complex plane, this paper gives another equivalent characterization of causal automorphisms in terms of M\"{a}rzke-Wheeler maps and proves for the first time that differentiable causal automorphisms are in fact conformal isometries. Moreover, we prove the following characterization of differentiable causal automorphisms in terms of Minkowski conformal maps:

\begin{teo}
In two dimensional Minkowski spacetime, $F$ is a $C^{1}$ causal automorphism if and only if $F$ is a $C^{1}$ Minkowski conformal map whose restriction to the time axis intersects every lightray.
\end{teo}

The above is a new result not included in the well known theorem by Hawking \cite{HKM}:

\begin{teo}
A causal isomorphism between strongly causal spacetimes of dimension strictly greater than two is a conformal isometry.
\end{teo}

In particular, the characterization of causal automorphisms in terms of M\"{a}rzke-Wheeler maps gives a negative answer to a recently raised question posed by Low \cite{Lo} and commented in \cite{Ki2}, who wonders whether two dimensional Minkowski $C^{2}$ causal automorphisms are characterized by their wave equation. However, we prove the following characterization for two dimensional Minkowski $C^{2}$ causal automorphisms:

\begin{teo}
$F$ is a $C^{2}$ causal automorphism if and only if $F$ is a $\mathcal{M}_{2}$-holomorphic or $\mathcal{M}_{2}$-antiholomorphic $C^{2}$ map (see Definition \ref{DefHolomorphic}) whose restriction to the time axis intersects every lightray.
\end{teo}

Finally, a proper time formula for accelerated observers in two dimensional Minkowski spacetime is given in the appendix. This formula solves the twin paradox in this space and reproduces the well-known slowing down of clocks in the gravitational acceleration direction.

\section{Preliminaries}

In what follows we will denote by $\mathcal{M}_{2}$ the two-dimensional Minkowski space and all causal morphisms $F$ are intended to be $F:\mathcal{M}_{2}\rightarrow \mathcal{M}_{2}$. The associated curve of a function $F:\mathcal{M}_{2}\rightarrow \mathcal{M}_{2}$ is the function restricted to the time axis; i.e. the function $\gamma$ such that $\gamma(s)=F(s,0)$ for every real $s$ (in general, $F$ will be continuous so $\gamma$ will be a continuous curve). Unless explicitly stated, differentiable maps are $C^{1}$ maps and continuity is relative to the Euclidean topology.

\begin{defi}
Consider a pair of points $x$ and $y$ in $\mathcal{M}_{2}$. We say that:
\begin{enumerate}
  \item $x$ causally precedes $y$ ($x<y$) if $x-y$ is a future directed null vector.
  \item $x$ chronologically precedes $y$ ($x<<y$) if $x-y$ is a future directed timelike vector.
\end{enumerate}
\end{defi}

\begin{figure}
\begin{center}
  \includegraphics[height=0.4\textwidth]{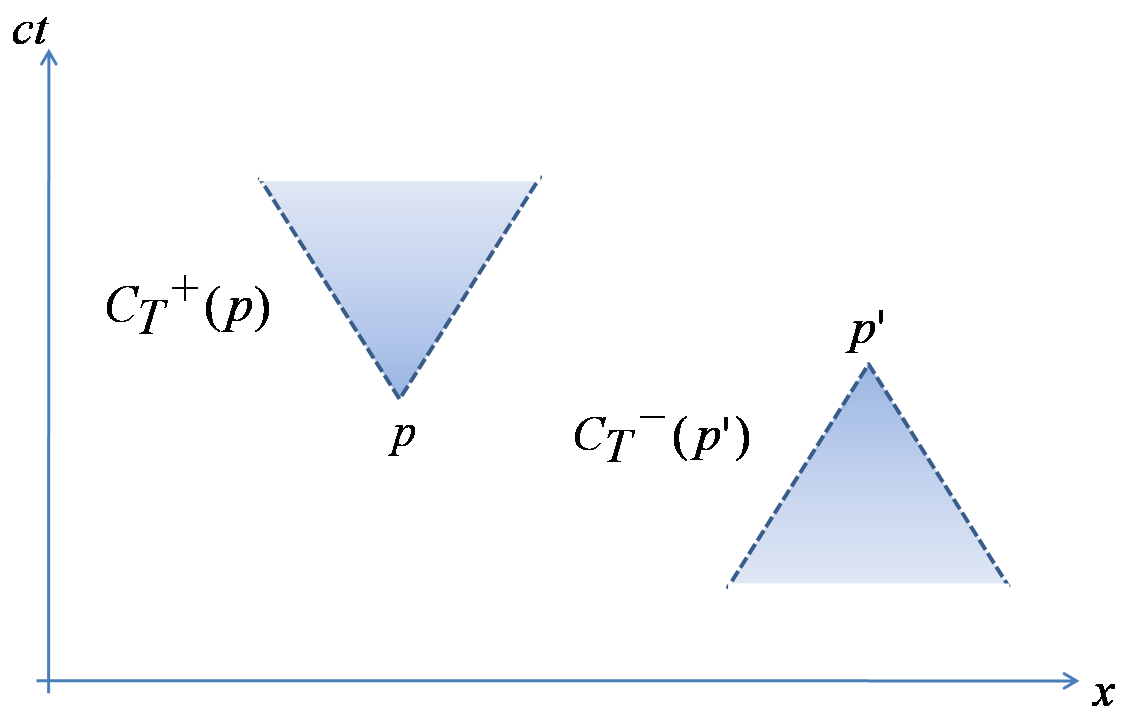}\\
  \end{center}
  \caption{Timelike future and past directed regions}\label{Conos}
\end{figure}
It will be convenient to define the following causal and chronological regions (see Figure \ref{Conos}):

\begin{itemize}
\item $\mathcal{C}_{N}^{+}(p)=\{x\in \mathcal{M}_{2}\ /\ p<x\}$
\item $\mathcal{C}_{N}^{-}(p)=\{x\in \mathcal{M}_{2}\ /\ x<p\}$
\item $\mathcal{C}_{T}^{+}(p)=\{x\in \mathcal{M}_{2}\ /\ p<<x\}$
\item $\mathcal{C}_{T}^{-}(p)=\{x\in \mathcal{M}_{2}\ /\ x<<p\}$
\item $\mathcal{C}_{N}(p)=\mathcal{C}_{N}^{+}(p)\cup \mathcal{C}_{N}^{-}(p)$
\item $\mathcal{C}_{T}(p)=\mathcal{C}_{T}^{+}(p)\cup \mathcal{C}_{T}^{-}(p)$
\end{itemize}

\begin{defi}\label{DefCausal}
$F$ is a causal morphism if $F$ preserves $<$; i.e. $x<y$ implies $F(x)<F(y)$. $F$ is a causal automorphism if $F$ is bijective and $F$ and $F^{-1}$ preserve $<$.
\end{defi}

In particular, if $F$ is a causal morphism then $F(l)\subset l'$ where $l$ and $l'$ are lightrays. Moreover, if $F$ is a causal automorphism then $F(l)=l'$ where $l$ and $l'$ are lightrays. For the following lemma see \cite{Na}.

\begin{lema}\label{equivalence}
$F$ is a causal automorphism if and only if $F$ is bijective and $F$ and $F^{-1}$ preserve $<<$.
\end{lema}

In two-dimensional Minkowski space $\mathcal{M}_{2}$ we can distinguish between right and left moving lightrays. Thus, for every point $p$ in $\mathcal{M}_{2}$ there exists a unique pair consisting of a left and a right moving lightrays $l_{L}(p)$ and $l_{R}(p)$ respectively such that $\{p\}=l_{L}(p)\cap l_{R}(p)$. By definition, it is clear that a causal automorphism maps parallel lightrays into parallel lightrays. This motivates the following definition:

\begin{defi}
A causal morphism $F$ is orientation preserving (reversing) if $F$ maps right moving lightrays into right (left) moving lightrays.
\end{defi}

We will see later that for $C^{1}$ causal automorphisms, the previous definition is equivalent to the differentiable one \cite{War}. It is important to remark that the concept of lightray in two dimensional spacetime is purely mathematical for electromagnetic theory in this dimension has no propagation. This is because there is no magnetic field so there are no electromagnetic waves. Moreover, the photon of two dimensional $QED$ is a free massive boson \cite{Sc}.

\section{Algebra of the two-dimensional Spacetime}

\begin{figure}
\begin{center}
  \includegraphics[height=0.35\textwidth]{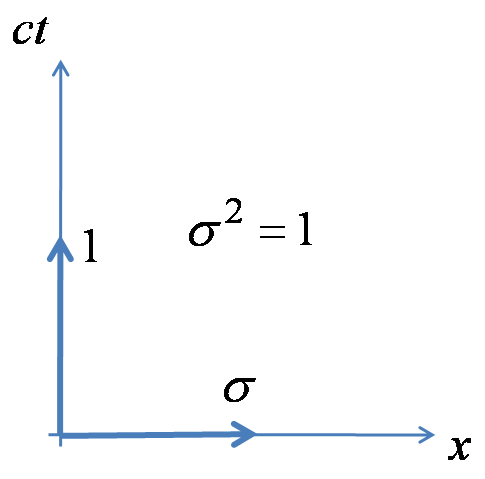}\\
  \end{center}
  \caption{Two dimensional Spacetime Algebra}\label{SpaceTimeAlgebra}
\end{figure}

Consider the associative real algebra $A=\R[\sigma]$ such that $\sigma^{2}=1$; i.e. $$A=\R[x]/\langle\ x^{2}-1\ \rangle$$ with the conjugation $\overline{a+b\ \sigma}=a-b\ \sigma$. Defining $|a|_{L}^{2}= \bar{a}\ a$ we have that $$|a\cdot b|_{L}^{2}=|a|_{L}^{2}\ |b|_{L}^{2}$$ This quadratic form comes from the inner product $\langle a,\ b \rangle= \Pi^{0}(\bar{a}\cdot b)$ where $\Pi^{0}$ is the projection over the first coordinate.
In what follows, we will identify the two dimensional spacetime $\mathcal{M}_{2}$ with the algebra $A$ through the correspondence (see Figure \ref{SpaceTimeAlgebra}) $$(x, c\ t)\leftrightarrow c\ t +x\ \sigma$$
where $c$ is the speed of light. It is interesting to see that this algebra encodes the usual special relativistic kinematic relations. The $2$-velocity $u$ associated to the $1$-velocity $v$ is $$u=\frac{c+v\ \sigma}{|c+v\ \sigma|_{L}}= \frac{1+\frac{v}{c}\sigma}{\sqrt{1-\frac{v^{2}}{c^{2}}}}$$
and the Lorentz transformation is just $p'=u\cdot p$; i.e. $$p'=u\cdot p= \frac{1+\frac{v}{c}\sigma}{\sqrt{1-\frac{v^{2}}{c^{2}}}}\cdot(ct+x\ \sigma)=c\ \frac{t+\frac{v}{c^{2}}x}{\sqrt{1-\frac{v^{2}}{c^{2}}}}+ \frac{x+vt}{\sqrt{1-\frac{v^{2}}{c^{2}}}}\ \sigma$$
The addition of velocities formula is just the product of the respective $2$-velocities: $$u\cdot u'=\frac{1+\frac{v}{c}\sigma}{\sqrt{1-\frac{v^{2}}{c^{2}}}}\cdot\frac{1+\frac{w}{c}\sigma}{\sqrt{1-\frac{w^{2}}{c^{2}}}}= \frac{1+\frac{v\ast w}{c}\sigma}{\sqrt{1-\frac{(v\ast w)^{2}}{c^{2}}}}$$ where $$v\ast w= \frac{v+w}{1+\frac{v\ w}{c^{2}}}$$
This way, Special Relativity in two dimensional spacetime becomes a Wick rotated version of the complex numbers.

\section{M\"{a}rzke-Wheeler Map}

\begin{figure}
\begin{center}
  \includegraphics[height=0.4\textwidth]{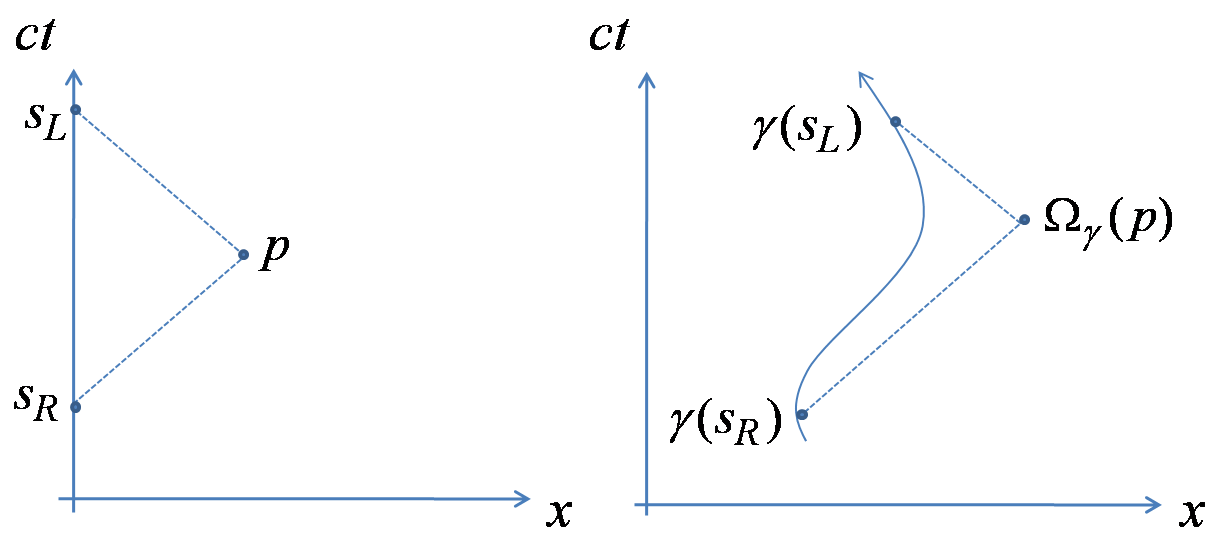}\\
  \end{center}
  \caption{M\"{a}rzke-Wheeler map}\label{MW_Coord}
\end{figure}

An observer is a continuous curve $\gamma: \R\rightarrow \mathcal{M}_{2}$ such that $\gamma(t)\in \mathcal{C}_{T}^{+}(\gamma(s))$ for every $t>s$ and $\gamma(t)\in \mathcal{C}_{T}^{-}(\gamma(s))$ for every $t<s$, for every real $s$. We define the M\"{a}rzke-Wheeler map $\Omega_{\gamma}:\mathcal{M}_{2}\rightarrow \mathcal{M}_{2}$ of an observer $\gamma$ as follows:
\begin{equation}
\{\Omega_{\gamma}(p)\}= l_{L}(\gamma(s_{L}))\cap l_{R}(\gamma(s_{R}))
\end{equation}
such that $\{(s_{L},0)\}=l_{L}(p)\cap \left(\R\times\{0\}\right)$ and $\{(s_{R},0)\}=l_{R}(p)\cap \left(\R\times\{0\}\right)$ (see Figure \ref{MW_Coord}). This map \cite{MW} is clearly an extension of the Einstein synchronization convention for non accelerated observers.

\begin{prop}\label{MWformula}
Consider an observer $\gamma$. Then,
\begin{equation}
\Omega_{\gamma}(s+x\sigma)= \frac{\gamma(s+x)+\gamma(s-x)}{2}+\frac{\gamma(s+x)-\gamma(s-x)}{2}\sigma
\end{equation}
\end{prop}
\dem
\begin{eqnarray*}
  |\Omega_{\gamma}(s+x\sigma)-\gamma(s\pm x)|_{L}^{2} &=& |\frac{\gamma(s+x)-\gamma(s-x)}{2}\cdot(\sigma\mp 1)|_{L}^{2} \\
   &=& |\frac{\gamma(s+x)-\gamma(s-x)}{2}|_{L}^{2}\ |(\sigma\mp 1)|_{L}^{2}=0
\end{eqnarray*}
because $|(\sigma\mp 1)|_{L}^{2}=0$.
\fdem

\begin{cor}\label{MWCont}
The M\"{a}rzke-Wheeler map of an observer is continuous.
\end{cor}

If $\gamma$ is a $C^{1}$ observer then the M\"{a}rzke-Wheeler map verifies the relation
\begin{equation}
\partial_{0}\Omega_{\gamma}= \sigma\ \partial_{1}\Omega_{\gamma}
\end{equation}
The above property implies that the M\"{a}rzke-Wheeler map has a wave like motion:
\begin{equation}
\Box\ \Omega_{\gamma}=0
\end{equation}
such that $\Omega_{\gamma}(s)= \gamma(s)$ for every real $s$ and $C^{2}$ observer $\gamma$. This motivates the following definition:

\begin{defi}\label{DefHolomorphic}
We say that a function $F:\mathcal{M}_{2}\rightarrow \mathcal{M}_{2}$ is $\mathcal{M}_{2}$-holomorphic if $\partial_{0}F= \sigma\ \partial_{1}F$ and $\mathcal{M}_{2}$-antiholomorphic if $\partial_{0}F= -\sigma\ \partial_{1}F$. If $F:\mathcal{M}_{2}\rightarrow \mathcal{M}_{2}$ is a $\mathcal{M}_{2}$-holomorphic function, we define its $\mathcal{M}_{2}$-derivative as $DF=\partial_{0}F= \sigma\ \partial_{1}F$. Is clear that $\mathcal{M}_{2}$-holomorphic and $\mathcal{M}_{2}$-antiholomorphic functions have wave like motion.
\end{defi}

\begin{prop}\label{MWConformal}
If $\gamma$ is a $C^{1}$ observer then $\Omega_{\gamma}$ is a conformal map and its conformal factor is $|D\Omega_{\gamma}|_{L}^{2}$.
\end{prop}
\dem
\begin{eqnarray*}
  \langle \partial_{i}\Omega_{\gamma},\ \partial_{j}\Omega_{\gamma} \rangle &=& \langle D\Omega_{\gamma}\ \varepsilon_{i},\ D\Omega_{\gamma}\ \varepsilon_{j} \rangle= \Pi^{0}( \overline{D\Omega_{\gamma}\ \varepsilon_{i}}\ D\Omega_{\gamma}\ \varepsilon_{j}) \\
   &=& \Pi^{0}( \overline{\varepsilon_{i}}\ \overline{D\Omega_{\gamma}}\ D\Omega_{\gamma}\ \varepsilon_{j})=|D\Omega_{\gamma}|_{L}^{2}\ \Pi^{0}( \overline{\varepsilon_{i}}\ \varepsilon_{j}) \\
   &=& |D\Omega_{\gamma}|_{L}^{2}\ \langle \varepsilon_{i},\ \varepsilon_{j} \rangle=|D\Omega_{\gamma}|_{L}^{2}\ \eta_{ij}
\end{eqnarray*}
where $\varepsilon_{0}=1$ and $\varepsilon_{1}=\sigma$.
\fdem

The above proposition shows that acceleration is equivalent to a conformal map in two dimensional flat spacetime. Because $R_{0101}=0$ where $R$ is the Riemann curvature tensor, we conclude the interesting fact that the conformal factor logarithm of the M\"{a}rzke-Wheeler map has also a wave like motion:
\begin{equation}
\Box\ ln\ g=0
\end{equation}
where $g=|D\Omega_{\gamma}|_{L}^{2}$ for $C^{2}$ observers.

\begin{lema}\label{MWCausal}
If $\gamma$ is an observer then $\Omega_{\gamma}$ is an orientation preserving continuous causal morphism.
\end{lema}
\dem
By Lemma \ref{MWformula} and the fact that $\gamma$ is continuous, is clear that $\Omega_{\gamma}$ is continuous. By definition, $\Omega_{\gamma}$ preserves $<$ and maps right (left) moving lightrays into right (left) moving lightrays.
\fdem

\section{$C^{1}$ Causal Automorphisms as Conformal Isometries}

\begin{defi}\label{Def}
We say an observer $\gamma:\R\rightarrow \mathcal{M}_{2}$ verifies the Lightray Intersecting Property (LIP) if every lightray intersects $\gamma(\R)$.
\end{defi}

\begin{lema}\label{causalCont}
If $F$ is a causal automorphism then its associated curve is an observer; i.e. $\gamma$ is an observer where $\gamma(s)=F(s)$ for every real $s$.
\end{lema}
\dem
To prove that $\gamma:\R\rightarrow \mathcal{M}_{2}$ is continuous it is enough to show that $\gamma$ maps a monotone convergent sequence into a convergent sequence. Consider a strictly ascending convergent sequence $(s_{n})$ of real numbers such that $s_{n}\rightarrow s_{0}$. Because of Lemma \ref{equivalence},
\begin{equation}
    \gamma(s_{1})<<\gamma(s_{2})<<\gamma(s_{3})\ldots <<\gamma(s_{0}) \label{RelCausal}
\end{equation}
and we have that $\gamma(s_{n})\in \overline{\mathcal{C}_{T}^{+}(\gamma(s_{1}))\cap \mathcal{C}_{T}^{-}(\gamma(s_{0}))}$ for every $n\geq 1$. In particular, the sequence $(\gamma(s_{n}))$ is contained in a compact set and by the Bolzano-Weierstrass Theorem, $(\gamma(s_{n}))$ has a limit point $p$ such that
\begin{equation}
    \gamma(s_{1})<<\gamma(s_{2})<<\gamma(s_{3})\ldots <<p \label{RelCausal2}
\end{equation}
For if there is a natural $N$ such that $\gamma(N)$ doesn't chronologically precede $p$ then $\overline{\mathcal{C}_{T}^{+}(\gamma(s_{N+1}))}\cap \overline{\mathcal{C}_{T}^{-}(p)}=\emptyset$ which implies (because of (\ref{RelCausal})) that $p$ is not the limit point of the subsequence $(\gamma(s_{n}))_{n>N}$ which is absurd. Consider an open disk $D_{\varepsilon}(p)$ centered at $p$ of radius $\varepsilon$. There is a natural $m$ such that $\gamma_{m}\in D_{\varepsilon}(p)$ and because of (\ref{RelCausal2}), for every $n>m$ we have $$\gamma(s_{n})\in\mathcal{C}_{T}^{+}(\gamma(s_{m}))\cap \mathcal{C}_{T}^{-}(p)\subset D_{\varepsilon}(p)$$
and we conclude that $(\gamma(s_{n}))\rightarrow p$. In particular, $$\overline{\mathcal{C}_{T}^{+}(p)}= \bigcap_{n\in \N} \mathcal{C}_{T}^{+}(\gamma(s_{n}))$$
If $p=\gamma(s_{0})$ we are done. If not, $\mathcal{C}_{T}^{+}(p)\cap \mathcal{C}_{T}^{-}(\gamma(s_{0}))$ is an open non empty set such that $$F^{-1}(\mathcal{C}_{T}^{+}(p)\cap \mathcal{C}_{T}^{-}(\gamma(s_{0})))\subset \left(\bigcap_{n\in \N} F^{-1}(\mathcal{C}_{T}^{+}(\gamma(s_{n})))\right)\cap F^{-1}(\mathcal{C}_{T}^{-}(\gamma(s_{0})))= $$ $$=\left(\bigcap_{n\in \N} (s_{n}, +\infty)\right)\cap (-\infty, s_{0})=[s_{0}, +\infty)\cap (-\infty, s_{0})=\emptyset$$
which is absurd. The argument for a descending sequence is analogous. We have shown that $\gamma$ is continuous. Is clear that the continuous curve $\gamma$ is an observer for $$\gamma(s+h)=F(s+h)>>F(s)=\gamma(s)$$ because $s+h>>s$ such that $h>0$ where $s$ and $h$ are real numbers.
\fdem

\begin{lema}\label{LIP}
Consider a causal morphism $F$. Then, $F$ is a causal automorphism if and only if its associated curve $\gamma$ is an observer verifying LIP.
\end{lema}
\dem
Suppose that $F$ is a causal automorphism. Because of Lemma \ref{causalCont}, $\gamma$ is an observer. Consider a lightray $l$. There is a unique lightray $l'$ such that $F(l')=l$ and intersects the real line (time axis) in the real $s_{0}$. Then, $l$ intersects $\gamma$ in the point $\gamma(s_{0})$. Conversely, consider a point $p$ in $\mathcal{M}_{2}$. There is a unique pair consisting of a left and a right moving lightrays $l_{L}(p)$ and $l_{R}(p)$ respectively such that $\{p\}=l_{L}(p)\cap l_{R}(p)$. Because $\gamma$ is an observer verifying LIP, $\{\gamma(s_{L})\}=l_{L}(p)\cap \gamma(\R)$ and $\{\gamma(s_{R})\}=l_{R}(p)\cap \gamma(\R)$. Because $F$ is a causal morphism, if $F$ is orientation preserving then $p=F(p')$ such that $\{p'\}=l_{L}(s_{L})\cap l_{R}(s_{R})$ for $$\{F(p')\}=F(l_{L}(s_{L})\cap l_{R}(s_{R}))= F(l_{L}(s_{L}))\cap F(l_{R}(s_{R}))\subset l_{L}(\gamma(s_{L}))\cap l_{R}(\gamma(s_{R}))=$$ $$= l_{L}(p)\cap l_{R}(p)= \{p\}$$ and $p'$ is the unique point verifying that property. If $F$ is orientation reversing then $p=F(p')$ such that $\{p'\}=l_{L}(s_{R})\cap l_{R}(s_{L})$.
\fdem

\begin{teo}\label{CausalimplicaMW}
$F:\mathcal{M}_{2}\rightarrow \mathcal{M}_{2}$ is a causal automorphism if and only if its associated curve $\gamma$ is an observer which verifies LIP and
\begin{enumerate}
  \item $F=\Omega_{\gamma}$ if $F$ is orientation preserving.
  \item $F=\Omega_{\gamma}\circ \bar{z}$ if $F$ is orientation reversing.
\end{enumerate}
where $\bar{z}$ is the conjugate map.
\end{teo}
\dem
Suppose that $F$ is orientation preserving. Because $F$ is a causal automorphism, by Lemmas \ref{causalCont} and \ref{LIP} $\gamma$ is an observer verifying LIP and then by Lemmas \ref{MWCausal} and \ref{LIP} $\Omega_{\gamma}$ is also a causal automorphism. By Definition \ref{DefCausal} and the remark below it, $F(l)= l'$ where $l$ and $l'$ are lightrays and this property is also verified by $\Omega_{\gamma}$. Because $\gamma(s)=F(s)$ for every real $s$ and $F$ is orientation preserving, then $$F(l)=\Omega_{\gamma}(l)$$ where $l$ is a lightray. Every point in $\mathcal{M}_{2}$ is a unique pair consisting of a left and a right lightray $l_{L}(p)$ and $l_{R}(p)$ respectively such that $\{p\}=l_{L}(p)\cap l_{R}(p)$. Then we have $$\{F(p)\}= F(l_{L}(p)\cap l_{R}(p))= F(l_{L}(p))\cap F(l_{R}(p))= \Omega_{\gamma}(l_{L}(p))\cap \Omega_{\gamma}(l_{R}(p))=$$ $$= \Omega_{\gamma}(l_{L}(p)\cap l_{R}(p))= \{\Omega_{\gamma}(p)\}$$ and we get the result. For the orientation reversing case just replace $\Omega_{\gamma}$ by $\Omega_{\gamma}\circ \bar{z}$ in the previous proof. Because of Lemmas \ref{MWCausal} and \ref{LIP}, the converse follows.
\fdem

The above characterization in terms of M\"{a}rzke-Wheeler maps agrees with the one given in \cite{Ki} and \cite{Lo}. In particular, we have shown that for $C^{1}$ causal automorphisms, definition \ref{Def} is equivalent to the usual notion of orientation preserving and reversing. Because of Lemmas \ref{MWCausal} and \ref{MWConformal}, the above theorem implies:

\begin{cor}
In two dimensional Minkowski space, every causal automorphism is continuous and every $C^{1}$ causal automorphism is a conformal isometry.
\end{cor}

\begin{prop}\label{C}
Let $F:\mathcal{M}_{2}\rightarrow \mathcal{M}_{2}$ be a conformal map such that its associated curve $\gamma$ is an observer verifying LIP. Then,
\begin{enumerate}
  \item $F=\Omega_{\gamma}$ if $F$ is orientation preserving.
  \item $F=\Omega_{\gamma}\circ \bar{z}$ if $F$ is orientation reversing.
\end{enumerate}
\end{prop}
\dem
Because $F$ is a conformal map, its differential $DF$ maps null vectors into null vectors and because $F$ is $C^{1}$, we conclude that $F(l)\subset l'$ where $l$ and $l'$ are lightrays. Suppose that $F$ is orientation preserving (in the usual sense). By Hypothesis, $\gamma$ is an observer verifying LIP and then by Lemmas \ref{MWCausal} and \ref{LIP}, $\Omega_{\gamma}$ is a causal automorphism. In particular, $\Omega_{\gamma}(l)= l'$ where $l$ and $l'$ are lightrays. Because $\gamma(s)=F(s)$ for every real $s$ and $F$ is orientation preserving, then $$F(l)\subset\Omega_{\gamma}(l)$$ where $l$ is a lightray. Every point in $\mathcal{M}_{2}$ is a unique pair consisting of a left and a right lightray $l_{L}(p)$ and $l_{R}(p)$ respectively such that $\{p\}=l_{L}(p)\cap l_{R}(p)$. Then we have $$\{F(p)\}= F(l_{L}(p)\cap l_{R}(p))= F(l_{L}(p))\cap F(l_{R}(p))\subset \Omega_{\gamma}(l_{L}(p))\cap \Omega_{\gamma}(l_{R}(p))=$$ $$= \Omega_{\gamma}(l_{L}(p)\cap l_{R}(p))= \{\Omega_{\gamma}(p)\}$$ and we get the result. For the orientation reversing case just replace $\Omega_{\gamma}$ by $\Omega_{\gamma}\circ \bar{z}$ in the previous proof.
\fdem

We have shown the following characterization theorem:

\begin{teo}
In two dimensional Minkowski spacetime, $F$ is a $C^{1}$ causal automorphism if and only if $F$ is a Minkowski conformal map whose associated curve $\gamma$ is an observer verifying LIP.
\end{teo}

\section{$C^{2}$ Causal Automorphisms as Minkowski (anti) Holomorphic Maps}

Recently, Low \cite{Lo} (page 4) has raised the question of whether causal automorphisms are characterized by wave equations:

\begin{flushleft}
\emph{``Comment: It is also worth observing that by considering the situation in terms of
Cartesian coordinates, we can see that X and T are both given by solutions of the wave
equation on $M^{2}$ (at least in the case where they are sufficiently differentiable). It would
be interesting to know whether there is a useful characterization of just which solutions
of the wave equation give rise to causal automorphisms of $M^{2}$."
}
\end{flushleft}

Paraphrased in our terms, Low asks if the solution of the problem $\Box F=0$ such that $\gamma$ is an observer verifying LIP where $\gamma(s)=F(s)$ for every real $s$, is necessarily a causal automorphism. Proposition \ref{CausalimplicaMW} gives a negative answer to that question, for a causal automorphism must be $\mathcal{M}_{2}$-holomorphic or $\mathcal{M}_{2}$-antiholomorphic and the general solution of the wave equation is a linear combination of both. For example, consider a pair of observers $\gamma_{1}$ and $\gamma_{2}$ and the function $F=\Omega_{\gamma_{1}}+\Omega_{\gamma_{2}}\circ \bar{z}$. Then, $F$ is a solution of the wave equation such that its associated observer is $\gamma_{1}+\gamma_{2}$. However, by the previous theorem, if $F$ is a causal automorphism then $F=\Omega_{\gamma_{1}+\gamma_{2}}$ or $F=\Omega_{\gamma_{1}+\gamma_{2}}\circ \bar{z}$, and we conclude that $F$ is not a causal automorphism.

However, we can give the following characterization for $C^{2}$ causal automorphisms:

\begin{lema}\label{unicidad}
If $F$ is a $\mathcal{M}_{2}$-holomorphic or $\mathcal{M}_{2}$-antiholomorphic $C^{2}$ function whose associated curve is zero then $F=0$.
\end{lema}
\dem
In this proof we forget the algebraic structure considered so far and treat $\mathcal{M}_{2}$ just as a real vector space. Suppose $F$ is $\mathcal{M}_{2}$-holomorphic and write $F(x,y)= (P(x,y),\ Q(x,y))$. Then, $$\partial_{x}P=\partial_{y}Q$$ $$\partial_{y}P=\partial_{x}Q$$ such that $P(0,y)=Q(0,y)=0$ for every real $y$. Because $P$ and $Q$ are $C^{2}$ real functions, the above equations and constraints are equivalent to the following: $$Q(x,y)=\int_{0}^{x}dx'\ \partial_{y}P(x',y)$$ such that $\Box P=0$, $P(0,y)=0$ and $\partial_{x}P(0,y)=0$ for every real $y$. These constraints imply that $P=0$ so $Q=0$ as well. We have proved that $F=0$. The $\mathcal{M}_{2}$-antiholomorphic case is similar.
\fdem

\begin{teo}
$F$ is a $C^{2}$ causal automorphism if and only if $F$ is a $\mathcal{M}_{2}$-holomorphic or $\mathcal{M}_{2}$-antiholomorphic $C^{2}$ function whose associated curve is an observer verifying LIP.
\end{teo}
\dem
The direct implication follows from proposition \ref{CausalimplicaMW}. For the converse, suppose that $F$ is a $\mathcal{M}_{2}$-holomorphic $C^{2}$ function whose associated curve is an observer $\gamma$. This way, $\gamma$ is $C^{2}$ and $\Omega_{\gamma}$ is also a $\mathcal{M}_{2}$-holomorphic $C^{2}$ function whose associated curve is the observer $\gamma$. Then, $F-\Omega_{\gamma}$ is a $\mathcal{M}_{2}$-holomorphic $C^{2}$ function whose associated curve is zero and by Lemma \ref{unicidad}, $F=\Omega_{\gamma}$. Lemmas \ref{MWCausal} and \ref{LIP} imply that $F$ is a causal automorphism. For the $\mathcal{M}_{2}$-antiholomorphic case just replace $\Omega_{\gamma}$ by $\Omega_{\gamma}\circ \bar{z}$ in the previous proof.
\fdem

Although a causal automorphism is not characterized by a wave equation, we have the following characterization in terms of it:
\numparts
\begin{cor}
Consider a $C^{2}$ observer $\gamma$ verifying LIP. $F$ is a $C^{2}$ causal automorphism whose associated observer is $\gamma$ if and only if
\begin{equation}
Q(x,y)= q(y)\pm\int_{0}^{x}dx'\ \partial_{y}P(x',y)
\end{equation}
such that $P$ verifies the following wave equation:
\begin{eqnarray}
  \Box\ P &=& 0 \\
  P(0,y) &=& p(y) \\
  \partial_{x}P(0,y) &=& \pm q'(y)
\end{eqnarray}
for every real $y$, where $\gamma(y)= (p(y), q(y))$ and $F(x,y)= (P(x,y), Q(x,y))$ for every pair of reals $x$ and $y$.
\end{cor}
\endnumparts

\appendix
\section{Proper time formula for Accelerated Observers in two dimensional Minkowski Spacetime}

In Special Relativity, the proper time of a given timelike continuous future directed curve $\alpha$ is
\begin{equation}
    \Delta\tau=\frac{1}{c}\int ds= \int\ \sqrt{1-\frac{v(t)^{2}}{c^{2}}}\ dt \label{ProperTimeFormula}
\end{equation}
where $v$ is $\alpha$'s $1$-velocity measured by an inertial observer $\gamma$. Following M\"{a}rzke-Wheeler synchronization convention for accelerated observers, by Lemma \ref{MWConformal} we have the following proper time formula relative to an accelerated observer $\gamma$: \begin{equation}
    \Delta\tau=\frac{1}{c}\int ds= \int\ |D\Omega_{\gamma}|_{L}(x(t), ct)\ \sqrt{1-\frac{v(t)^{2}}{c^{2}}}\ dt \label{ProperTimeAccFormula}
\end{equation}
where $v(t)$ is $\alpha$'s $1$-velocity and $x(t)$ is $\alpha$'s position measured by the accelerated observer $\gamma$ at the instant $t$. Formula \ref{ProperTimeAccFormula} simplifies to \ref{ProperTimeFormula} if the observer is an inertial one for the conformal factor is one in this case. This way, formula \ref{ProperTimeAccFormula} is a generalization of \ref{ProperTimeFormula}.

The twin paradox is the following: Consider a couple of twins $A$ and $B$ that use formula \ref{ProperTimeFormula} to calculate the brother's proper time. The twin $B$ launches in an accelerated space shuttle and then come back, relative to an inertial observer $A$ who stays in the air base, and finds that $A$ is younger than him at the same time that $A$ also finds that $B$ is younger than him, so, Who is the younger one? The abuse made by considering that certain formulas deduced for inertial observers remain valid for non-inertial ones is the origin of the twin paradox. Formula \ref{ProperTimeAccFormula} solves the twin paradox for the accelerated twin $B$ calculates the proper time of the inertial one $A$ with formula \ref{ProperTimeAccFormula} instead of using \ref{ProperTimeFormula}. This way both twins agree that $B$ is the younger one.

Without an explicit proper time formula, a similar approach to the twin paradox is also studied in \cite{PV}. Because of the fact that
\begin{equation}
\Box\ ln\ g=0
\end{equation}
where $g=|D\Omega_{\gamma}|_{L}^{2}$ for $C^{2}$ observers, we have shown that the proper time formula for an observer in two dimensional flat spacetime is:
\numparts
\begin{equation}
    \Delta\tau= \int\ e^{h(x(t), ct)}\ \sqrt{1-\frac{v(t)^{2}}{c^{2}}}\ dt \label{ProperTimeAccGenFormula}
\end{equation}
where $h$ is a scalar field such that
\begin{equation}
\Box\ h=0
\end{equation}
\endnumparts
As an application of formula \ref{ProperTimeAccFormula} consider a uniformly accelerated observer:
\numparts
\begin{equation}
\gamma(s)= \frac{c^{2}}{a}\ \exp\left(\frac{a\ s}{c^{2}}\ \sigma\right)\ \sigma
\end{equation}
where $a$ denotes its acceleration and $c$ the lightspeed. By proposition \ref{MWformula}, the M\"{a}rzke-Wheeler map $\Omega_{\gamma}$ of the observer $\gamma$ is:
\begin{equation}
\Omega_{\gamma}(z)= \frac{c^{2}}{a}\ \exp\left(\frac{a\ z}{c^{2}}\ \sigma\right)\ \sigma
\end{equation}\endnumparts
The conformal factor is:
\begin{equation}
|D\Omega_{\gamma}|_{L}(s + x\sigma)=\exp\left(\frac{a\ x}{c^{2}}\right)
\end{equation}
By the equivalence principle, we can think that the observer is at rest in a constant gravitational field with gravitational acceleration $g=-a$. By formula \ref{ProperTimeAccFormula} we have that:
\begin{equation}
\Delta\tau(x)= \exp\left(-\frac{g\ x}{c^{2}}\right)\ \Delta t
\end{equation}
where $\Delta\tau(x)$ is the time interval measured at $x$ by the observer $\gamma$. This way we have the formula:
\begin{equation}
\Delta\tau(x_{2})= \exp\left(-\frac{g\ \Delta x}{c^{2}}\right)\ \Delta\tau(x_{1})
\end{equation}
which express the well-known slowing down of clocks in the gravitational acceleration direction \cite{Wa}.

\end{document}